\documentstyle[12pt]{article}

\begin{document}

\mbox{} \hskip 10cm DF/IST-9.93

\mbox{} \hskip 10cm UATP-93/05

\mbox{} \hskip 10cm August 1993

\vskip 2cm

\begin{center}
THE BLACK HOLES OF A GENERAL TWO-DIMENSIONAL DILATON
GRAVITY THEORY \\
\vskip 1cm
{\bf Jos\'e P. S. Lemos} \\
\vskip 0.3cm
{\scriptsize  Departamento de Astrof\'{\i}sica,
              Observat\' orio Nacional-CNPq,} \\
{\scriptsize  Rua General Jos\'e Cristino 77,
              20921 Rio de Janeiro, Brazil,} \\
{\scriptsize  \&} \\
{\scriptsize  Departamento de F\'{\i}sica,
              Instituto Superior T\'ecnico,} \\
{\scriptsize  Av. Rovisco Pais 1, 1096 Lisboa, Portugal.} \\

\vskip 0.6cm
{\bf Paulo M. S\'a} \\
\vskip 0.3cm
{\scriptsize Unidade de Ci\^encias Exactas e Humanas,
             Universidade do Algarve,} \\
{\scriptsize Campus de Gambelas,
             8000 Faro, Portugal.}
\end{center}

\bigskip

\begin{abstract}
\noindent
A general dilaton gravity theory in 1+1 spacetime dimensions
with a cosmological constant $\lambda$ and a new dimensionless
parameter $\omega$,
contains as special cases the constant curvature theory
of Teitelboim and Jackiw, the theory equivalent to vacuum
planar General Relativity,
the first order string theory, and a two-dimensional purely
geometrical theory.
The equations of this general two-dimensional theory admit
several different black holes with various types of singularities.
The singularities can be spacelike, timelike or null,
and there are even cases without singularities.
Evaluation of the ADM mass, as a charge density integral,
is possible in some situations,
by carefully subtrating the black hole solution
from the corresponding linear dilaton at infinity.
\end{abstract}

\newpage

\noindent
{\bf 1. Introduction}

\vskip 3mm

\noindent
Black holes and spacetime singularities are fundamental
concepts which have been taken into account in the search
for the possible links between General Relativity
and Quantum Mechanics.
The black hole concept is connected to the non-local notion
of an event horizon.
On the other hand, a spacetime singularity is a concept
which is usually associated with unbound values of several
physical quantities.
Both concepts are not related a priori.
Yet, in classical General Relativity,
the existence of a black hole implies the existence of
a spacetime singularity.
It is conjectured that quantum effects of spacetime will
play a role in explaining the singula\-rity as a true
quantum object.
A possible quantum mechanical framework for
General Relativity is provided by string theory.
Within this theory, black hole solutions have been found for
various number of dimensions \cite{gidd}.
These solutions in turn,
allow for the possibility of analysing the
singula\-rity problem from new perspectives.
In particular,
black hole solutions in two-dimensional (2D) spacetimes
have been playing an important role in the understanding
of these issues.
Indeed, in pertubation theory,
the string field equations yield a 2D
black hole with spacetime singularities similar to the
Schwarzschild metric \cite{mand,witt}.
In contrast,
in the exact conformal quantum field theory \cite{witt,dijk}
it has been shown that the corresponding spacetime is
free of singularities \cite{perr}.
This is welcome since it provides an example where the exact
quantum description prevents the formation of a singularity.
There are several 2D theories with non-trivial dynamics.
One that has been widely studied is the constant curvature
theory of Teitelboim and Jackiw \cite{teit,jack}.
Within this theory one can show that the corresponding black
hole is also free of spacetime singularities and with a
structure analogous to the exact string solution \cite{sa}.
A 2D theory, which also has been discussed,
is the $R=T$ theory \cite{mann,lesa}.
Another 2D theory, introduced in \cite{lemo},
equivalent to planar General Relativity,
also admits black hole solutions which,
as expected, give rise to spacetime singularities.

In order to see how these features,
such as existence or not of singularities and event horizons,
might develop from theory to theory,
it is important to have a general theory
in which the above 2D theories could be connected,
each being a special case of the general theory.
In this manner,
results found in one of the theories could be related
straightforwardly with the others.
Brans-Dicke theory in 2D provides such a link \cite{lemo}.
The theory is specificied by two fields,
the dilaton $\phi$ and the graviton $g_{\mu \nu}$,
and two parameters, the cosmological constant $\lambda$ and
the parameter $\omega$.
In this paper we solve and analyse black hole and related
solutions for any values of the parameters
$\omega$ and $\lambda$.
In order to distinguish black hole solutions from other solutions
we have to define properly the notion of black hole event horizon.
As it is known \cite{hawk} one has first to characterize the
possibility of escaping to futureinfinity, which we denote
by ${\cal J}^+$
(although it does not necessarily mean future null infinity).
We then say that the spacetime $M$ contains a black hole  if
$M$ is not completely contained in the causal past of future
infinity denoted by $J^- ({\cal J}^+)$.
The black hole region is then given by $M-J^- ({\cal J}^+)$,
whose boundary is defined as being the event horizon.
In section~2 we set the equations.
In section~3 we divide the solutions in different classes,
depending on the values of the various parameters,
and study their causal structure with the
corresponding Penrose diagram.
In section~4 we calculate the ADM masses of the solutions.
Finally, in section~5 we conclude and comment on the
links between the solutions.

\vskip 1cm

\noindent
{\bf 2. The Equations}

\vskip 3mm

\noindent
We propose to solve the following action,
\begin{equation}
S=\frac{1}{2\pi} \int d^2x \sqrt{-g} e^{-2\phi}
  \left( R - 4 \omega \left( \partial \phi \right)^2
           + 4 \lambda^2
  \right),  \label{eq:1}
\end{equation}
where
$g$ is the determinant of the 2D metric,
$R$ is the curvature scalar,
$\phi$ is a scalar field,
$\lambda$ is a constant
and $\omega$ is a parameter.
For $\omega=0$ one obtains the Teitelboim-Jackiw theory,
for $\omega=-{1\over2}$ one has planar General Relativity,
for $\omega=-1$ one has the first order string theory
and for $\omega=\pm\infty$ one obtains a pure geometrical
theory.
Equation (\ref{eq:1}) is a Brans-Dicke type action
in two-dimensions.
A generalized action, where $\phi=\phi(\psi)$,
$\omega=\omega(\psi)$ and $\lambda=\lambda(\psi)$
as also been proposed \cite{bank}.
Here we treat $\omega$ and $\lambda$ as constants.

Varying this equation with respect to $g^{ab}$ and $\phi$
one gets the following equations,
\begin{eqnarray}
   & &  e^{2\phi} T_{ab} \equiv
       -2\left(\omega+1\right) D_a\phi D_b\phi
       +D_aD_b\phi
       -g_{ab} D_cD^c\phi   \nonumber \\
   & &  \ \ \ \ \ \ \ \ \ \ \ \ \ \ \ \ \ \ \ \ \
       +\left(\omega+2\right) g_{ab} D_c\phi D^c\phi
       -g_{ab}\lambda^2=0,  \label{eq:2}  \\
   & &  R -4\omega D_cD^c\phi
          +4\omega D_c\phi D^c\phi
          +4\lambda^2 =0,      \label{eq:3}
\end{eqnarray}
where $D$ represents the covariant derivative.

In order to find black hole solutions we write the metric
in the unitary gauge,
\begin{equation}
     ds^2 = - e^{2\nu} dt^2 + dx^2,   \label{eq:4}
\end{equation}
where $\nu$ is a function of $x$ and $t$.
If we assume the solution to be static in these coordinates,
then equations (2) and (3) reduce to the
following three equations:
\begin{eqnarray}
  & &  \phi_{,xx}
      -\left(\omega+2\right)\phi_{,x}^{ \ 2}
      +\lambda^2 = 0,               \label{eq:5}  \\
  & &  \omega\phi_{,x}^{ \ 2}
      +\phi_{,x}\nu_{,x}
      +\lambda^2 = 0,               \label{eq:6}  \\
  & &  2\omega\phi_{,xx}
      +\nu_{,xx}
      -2\omega\phi_{,x}^{ \ 2}
      +2\omega\phi_{,x}\nu_{,x}
      +\nu_{,x}^{ \ 2}
      -2\lambda^2 = 0.              \label{eq:7}
\end{eqnarray}
Note that the Bianchi identities in vacuum imply that
2D spacetimes have only two independent equations;
thus we can consider (\ref{eq:7}) as a spurious equation.

\vskip 1cm

\noindent
{\bf 3. The solutions}

\vskip 3mm

\noindent
Equation (\ref{eq:5}) can be cast in a more elucidative form.
If one defines $\Phi=e^{-(\omega+2)\phi}$, then (\ref{eq:5})
turns into $\Phi_{,xx}=(\omega+2)\lambda^2\Phi$.
Thus, the character of the solution
(hyperbolic, trigonometric or linear)
will depend on the sign of $(\omega+2)\lambda^2$.
We consider the three cases separately:
$(\omega+2)\lambda^2>0$,
$(\omega+2)\lambda^2<0$ and
$(\omega+2)\lambda^2=0$.

\vskip 0.8cm
\noindent
{\bf 3.1. $(\omega+2)\lambda^2>0$}

\noindent
In this case the general solution of eq.~(\ref{eq:5}) is
\begin{equation}
\phi= - \frac{1}{\omega+2} \ln
        \left[ A \cosh \left(
                 \sqrt{(\omega+2)\lambda^2} x \right)
              +B \sinh \left(
                 \sqrt{(\omega+2)\lambda^2} x \right)
        \right],                  \label{eq:8}
\end{equation}
where $A$ and $B$ are constants of integration.
Without loss of generality, this solution can be subdivided
in three different classes, $A>|B|$, $|A|<|B|$ and $A=|B|$.

\vskip 0.3cm
\noindent
{\bf 3.1.1. $A>|B|$}

\noindent
As we will show this case yields eight black holes.
Whenever  $A>|B|$,
the solution (\ref{eq:8}) can be written as
\begin{equation}
  \phi = -\frac{1}{\omega+2} \ln
               \cosh \left( \sqrt{(\omega+2) \lambda^2} x \right)
         +\phi_0,          \label{eq:9}
\end{equation}
which is defined for $-\infty<x<\infty$.
Inserting this solution in eq.~(\ref{eq:6}) we obtain
the metric:
\begin{equation}
  ds^2 = - \tanh^2 \left( \sqrt{(\omega+2)\lambda^2} x \right)
           \cosh^{ \frac{4\left(\omega+1\right)}{\omega+2}}
                   \left( \sqrt{(\omega+2)\lambda^2} x \right)
           dt^2
         + dx^2,                    \label{eq:10}
\end{equation}
where the third constant of integration was absorved by a time
rescaling.
To clarify some features of the metric it is worth to write
it in the Schwarz\-schild gauge,
by defining the radial coordinate $r$,
\begin{equation}
  r=\frac{b^{\frac{\omega+1}{\omega+2}}}{a}
            \cosh^{\frac{2(\omega+1)}{\omega+2}}
                 \left(\sqrt{(\omega+2)\lambda^2} x \right),
                 \label{eq:11}
\end{equation}
with
\begin{equation}
   a=\frac{2(\omega+1) \lambda^2}{\sqrt{(\omega+2)\lambda^2}},
                     \label{eq:12}
\end{equation}
and $b=\mbox{constant}>0$.
We draw attention to the fact, that the line
$-\infty<x<+\infty$ corresponds to the segment
$1<ar/(b^{\frac{\omega+1}{\omega+2}})<+\infty$;
each pair of space inverted points degenerates into just one
$r$.
Due to this important circunstance true event horizons will form
in cases where black holes were not expected, as it will become
clear.
Using eq.~(\ref{eq:11})
the dilaton and metric fields take the form,
\begin{eqnarray}
   & & \phi= - \ln (ar)^{\frac{1}{2 \left( \omega+1 \right)}},
                \label{eq:13}  \\
   & & ds^2 = - \left[ a^2 r^2
              - b (ar)^{\frac{\omega}{\omega+1}} \right]
                dt^2
              + \frac{dr^2}{a^2 r^2
              - b (ar)^{\frac{\omega}{\omega+1}}},
                   \label{eq:14}
\end{eqnarray}
where in eq.~(\ref{eq:13}) we have set to zero the constant of
integration.
In section~4 we show that the constant $b$ (or $\phi_0$)
is related with the ADM mass of the solution.
Note that $ar=b^{\frac{\omega+1}{\omega+2}}$
gives the radius of the horizon and also that in the unitary
gauge the value of the dilaton at the horizon,
$x=0$, is $\phi=\phi_0$.

In the Schwarzschild gauge the solutions are expressed
as power-laws in the radial variable $ar$.
The cases $\omega=-1$ and $\omega=-2$ seem to have
to be handled separately,
but in fact one can treat them as limiting cases.
As usual, when one has power-law solutions, one also
expects exponential and logarithmic solutions.
These are precisely given by the cases
$\omega=-1$ and $\omega=-2$, respectively.

An important quantity,
which signals the appearance of a singularity,
is the scalar curvature,
given by
\begin{equation}
  R=-2\lambda^2 \left[
         \frac{4\left(\omega+1\right)^2}{\omega+2}
        +\frac{2\omega b}{\omega+2}
         (ar)^{-\frac{\omega+2}{\omega+1}}
                \right].
                     \label{eq:15}
\end{equation}

Since for different values of $b$ the
causal structure does not change,
we take here $b=1$
(which as we shall see in section~4 is
equivalent to $\phi_0=0$).
To find the maximal analytical extension of the metric
one has to write it in the conformal gauge.
The conformal radial coordinate $r_*$ is found from
\begin{equation}
    r_*=\int  \frac{dr}{a^2r^2 - (ar)^{\frac{\omega}{\omega+1}}}.
                       \label{eq:16}
\end{equation}
In the advanced and retarded null coordinates,
\begin{equation}
  u=t-r_*,\quad\quad v=t+r_*,  \label{eq:17}
\end{equation}
the metric becomes:
\begin{equation}
  ds^2 = - \left[ a^2r^2
                 -(ar)^{\frac{\omega}{\omega+1}} \right] dudv.
                       \label{eq:18}
\end{equation}
The maximal analytical extension is found through the Kruskal
coordinates, given by
\begin{eqnarray}
  & & U=-\frac{1}{\sqrt{(\omega+2)\lambda^2}}
        e^{-\sqrt{(\omega+2)\lambda^2}u},   \label{eq:19} \\
  & & V= \frac{1}{\sqrt{(\omega+2)\lambda^2}}
        e^{ \sqrt{(\omega+2)\lambda^2}v}.   \label{eq:20}
\end{eqnarray}

In general, the integral in eq.~(\ref{eq:16}) does not have an
analytical expression.
Moreover,
the maximal analytical extension depends critically
on the values of $\omega$.
In what follows we consider rational values of $\omega$.
There are seven cases which have to be treated separately:
$0<\omega<+\infty$,
$\omega=0$,
$-1<\omega<0$,
$\omega=-1$,
$-2<\omega<-1$,
$-\infty<\omega<-2$ and
$\omega=\mp\infty$.

\vskip 0.3cm
\noindent
{\bf 3.1.1.1.  $0<\omega<+\infty$}

\noindent

Within this range of values of $\omega$ there is no general
analytical solution for the integral in eq.~(\ref{eq:16}).
Thus, one should analyse a typical case.
Here we consider $\omega=1$.
The dilaton and the metric in Schwarzschild coordinates are
\begin{eqnarray}
  & &  e^{-2\phi}=\sqrt{ar}, \label{eq:21} \\
  & &  ds^2 = - \left( a^2 r^2 - \sqrt{ar} \right) dt^2
              + \frac{dr^2}{a^2 r^2 - \sqrt{ar}}, \label{eq:22}
\end{eqnarray}
with $a=\frac{4|\lambda|}{\sqrt3}$.
The conformal radial coordinate can be found
from eq.~(\ref{eq:16}),
\begin{equation}
   r_*=\frac{1}{3a} \ln \frac{(\sqrt{ar}-1)^2}{ar+\sqrt{ar}+1}
      -\frac{2\sqrt3}{3a} \arctan \frac{2\sqrt ar+1}{\sqrt3}.
                 \label{eq:23}
\end{equation}
In the Kruskal coordinates (\ref{eq:19}) and (\ref{eq:20})
the metric takes the form
\begin{equation}
    ds^2 = - \frac{\sqrt{ar}-1}{|\sqrt{ar}-1|}
             \sqrt{ar} \left( ar+\sqrt{ar}+1 \right)^{\frac32}
             e^{\sqrt{3} \arctan \frac{2\sqrt{ar}+1}{\sqrt3}}
             dUdV,       \label{eq:24}
\end{equation}
and
\begin{equation}
   UV = - \frac{1}{3\lambda^2}
          \frac{\sqrt{ar}-1}{\sqrt{ar+\sqrt{ar}+1}}
          e^{-\sqrt3\arctan \frac{2\sqrt{ar}+1}{\sqrt3}}.
                       \label{eq:25}
\end{equation}
One can now draw the Penrose diagram.
For $ar\rightarrow +\infty$, one has
$UV=-\frac{1}{3\lambda^2}e^{-\frac{\sqrt3\pi}{2}}$,
which is a vertical hiperbola in the Kruskal picture,
i.e.,
a straight vertical (timelike)
line in the Penrose diagram (see Fig.~1).
There is a horizon at $ar=1$, $UV=0$.
We recall that this region, $1<ar<\infty$, is degenerate
in the sense that each point $r$ corresponds to a pair
($-x,x$) in the unitary gauge.
Thus, region I is formed by two distinct triangles glued at
$ar=1$ ($x=0$).
Note now,
that in Kruskal coordinates the metric is not analytic at $ar=1$.
{}From eq.~(\ref{eq:24}) we see that on passing
from $ar>1$ to $ar<1$,
the metric changes sign.
Thus in region II, for $ar<1$,
one has to do $U\rightarrow \bar{U}=-U(-u)$
in order that the time direction is vertical.
In this patch one has
$\bar{U}V=-\frac{1}{3\lambda^2}e^{-\frac{\pi}{2\sqrt3}}$
for $ar=0$.
Thus the singularity at $ar=0$ is also a vertical hyperbola,
i.e., is a timelike singularity.
To regions I and II there would correspond regions
I' and II', obtained through the transformation
$u\rightarrow -u$, $v\rightarrow -v$,
where the light cones are reversed.
However,
no trajectory can leave regions I and II to enter regions
I' and II', and vice-versa.
For this reason,
regions I and II are totally disconnected from I' and II'.
The latter regions are simply a replic of the former,
and therefore there is no need to draw them.
The future infinity, ${\cal J}^+$, can be properly defined at
$ar\rightarrow \infty$.
As a matter of fact, there are two disconnected
${\cal J}^+$ at $x=\pm\infty$.
Observers in these two disconnected regions can only
communicate by passing through the horizon
$ar=1$ ($x=0$) into region II.
The solution is indeed a black hole, because the
causal past $J^-({\cal J}^+)$ is not the whole diagram.
One can now multiply region I and II {\it ad infinitum},
obtaining a Penrose diagram which is similar to the one
corresponding to the extreme 4-dimensional
Reissner-Nordstrom black hole solution.
Note, however,
that whereas inside the horizon we have
here a time-dependent metric,
in the extreme Reissner-Nordstrom case the metric is static.
The scalar curvature,
$R=-\frac{4\lambda^2}{3} \left( 8+\frac{1}{(ar)^{3/2}} \right)$,
is negative for all $r$,
which means that gravity has a non-attractive
character everywhere.
In particular,
it gives a timelike singularity inside the horizon.

\vskip 0.3cm
\noindent
{\bf 3.1.1.2.  $\omega=0$}

\noindent
This case gives the Teitelboim-Jackiw theory, where the scalar
curvature is constant, $R=-4\lambda^2$.
We have shown \cite{sa} that, surprisingly, this constant
curvature theory has two important features:
first, it admits a black hole, and second, the black hole
is free of spacetime singularities.
The maximal analytical extension gives a chain of universes
connected by timelike wormholes (see fig.~2 for Penrose diagram).
In Schwarzschild coordinates the dilaton
and the metric take the form
\begin{eqnarray}
  & & e^{-2\phi}=ar,    \label{eq:26} \\
  & & ds^2 = - \left( a^2r^2-1 \right) dt^2
             + \frac{dr^2}{a^2r^2-1},
                               \label{eq:27}
\end{eqnarray}
where $a=\sqrt2|\lambda|$.
There is again a duplication of region I, so that observers at each
end of the line $x\rightarrow\pm\infty$ can only communicate
if they enter through $x=0$ ($ar=1$) into region II.
An identical duplication applies to region III.
The non-singular character of the whole spacetime is analogous
to the behavior of the exact string black hole \cite{perr}.
The primed regions (I', II' and III') are copies of the unprimed
ones.

\vskip 0.3cm
\noindent
{\bf 3.1.1.3. $-1<\omega<0$}

\noindent
For this range of values of $\omega$ there is one which
is particulary important, $\omega=-\frac12$.
This gives planar General Relativity \cite{lemo}.
The dilaton and metric in Schwarzschild coordinates are given by
\begin{eqnarray}
   & & e^{-2\phi} = a^2r^2,  \label{eq:28} \\
   & & ds^2 = - \left( a^2r^2 - \frac{1}{ar} \right) dt^2
              + \frac{dr^2}{a^2r^2 - \frac{1}{ar}}, \label{eq:29}
\end{eqnarray}
where $a=\sqrt{\frac23} |\lambda|$.
The scalar curvature is
$R=-\frac{4\lambda^2}{3} \left( 1-\frac{1}{(ar)^3} \right)$.
There is a spacelike singularity at $ar=0$, where $R=+\infty$.
Inside the horizon the curvature is positive;
it passes through zero at the horizon, $ar=1$, and then becomes
negative.
Infinity is represented by a timelike line,
as in the four-dimensional anti-de~Sitter spacetime.
The Penrose diagram is given in figure~3.
Region I can also be duplicated by using the unitary gauge.
However, we note that, contrary to the two previous cases,
the black hole character of the solution does not depend
on this duplication.
In other words, solution (\ref{eq:28})-(\ref{eq:29})
is by itself a black hole solution.
This remark also applies to the black holes described
in the following three subsections (3.1.1.4.-3.1.1.6).

\vskip 0.3cm
\noindent
{\bf 3.1.1.4. $\omega=-1$}

\noindent

This is the original two-dimensional black hole \cite{mand,witt};
it gives the exponential metric.
In the Euclideanized version is called the cigar space.
{}From eq.~(\ref{eq:12}) we see that for $\omega=-1$ we loose
the scale, $a=0$.
Then, we have to define a new coordinate,
$r\rightarrow \frac{1}{a} \ln ar$.
The dilaton and metric are then:
\begin{eqnarray}
   & & e^{-2\phi}=e^{2|\lambda| r}, \label{eq:30} \\
   & & ds^2 = - \left( 1- e^{-2 |\lambda r|} \right) dt^2
              + \frac{dr^2}{1-e^{-2 |\lambda r|}}. \label{eq:31}
\end{eqnarray}
The range of $r$ is $-\infty<r<+\infty$.
The scalar curvature is $R=4\lambda^2 e^{-2|\lambda| r}$,
and so there is a singularity at $r\rightarrow -\infty$.
Spacetime is asymptotically flat for $r\rightarrow +\infty$.
The Penrose diagram is shown in figure~4;
it is identical to the Schwarzschild black hole.
Two new regions above and below the singularity have
been discussed in literature \cite{witt,dijk}.

\vskip 0.3cm
\noindent
{\bf 3.1.1.5. $-2<\omega<-1$}

\noindent
Here we analyse two typical cases, $\omega=-\frac32$ and
$\omega=-\frac43$, which give
different causal structures.

For $\omega=-\frac32$ the dilaton and metric are
\begin{eqnarray}
   & & e^{-2\phi}=\frac{1}{a^2 r^2}, \label{eq:32}  \\
   & & ds^2 = - \left( a^2r^2 - a^3r^3 \right) dt^2
              + \frac{dr^2}{a^2r^2-a^3r^3},  \label{eq:33}
\end{eqnarray}
with $a=-\sqrt2|\lambda|$.
In Schwarzschild coordinates the scalar curvature is given by
$R=-4\lambda^2\ (1-3ar)$.
There is a spacelike singularity at $ar\rightarrow +\infty$
and a timelike singularity at $ar\rightarrow -\infty$.
It is useful to change the radial coordinate into
$az={1\over ar}$.
The metric then takes the form,
\begin{equation}
ds^2 = \frac{1}{a^2 z^2}
       \left[ - \left( 1- \frac{1}{az} \right) dt^2
              + \frac{dz^2}{1-\frac{1}{az}} \right],
                                 \label{eq:34}
\end{equation}
We see that in these coordinates the metric is conformal
to the Schwarz\-schild metric.
However,
at $ar=0$ ($az=+\infty$) one can continue the solution
into the other singularity at $ar=-\infty$.
To make the manifold spatially complete one extends it
vertically, giving
an infinite chain of universes
(see Fig.~5).

For $\omega=-\frac43$ the dilaton and metric are
\begin{eqnarray}
   & & e^{-2\phi}=\frac{1}{a^3 r^3}, \label{eq:34a}  \\
   & & ds^2 = - \left( a^2r^2 - a^4r^4 \right) dt^2
              + \frac{dr^2}{a^2r^2-a^4r^4},  \label{eq:34b}
\end{eqnarray}
with $a=-\sqrt\frac23 |\lambda|$ and
$R=-\frac43 \lambda^2\ (1-6(ar)^2)$.
Now, the singularities at $ar=\pm\infty$ are both spacelike.
The maximal analytical extension is given by a diagram
that covers the whole plane (see Fig.~6).
These two diagrams cover all possibilities for rational
$\omega$.

\vskip 0.3cm
\noindent
{\bf 3.1.1.6. $-\infty<\omega<-2$}

\noindent
As a typical case we take $\omega=-3$.
Note that since $\omega+2<0$ we are considering $\lambda^2<0$.
The dilaton and the metric in Schwarzschild
coordinates take the form:
\begin{eqnarray}
   & & e^{-2\phi}=\frac{1}{\sqrt{ar}}, \label{eq:35}  \\
   & & ds^2 = - \left[ a^2r^2 - (ar)^{3/2} \right] dt^2
              + \frac{dr^2}{a^2r^2 - (ar)^{3/2}},
        \label{eq:36}
\end{eqnarray}
where $a=4|\lambda|$.
The scalar curvature is
$R=4\lambda^2 \left( 8-\frac{3}{\sqrt{ar}} \right)$,
which blows up $ar=0$.
The singularity is a null line hidden inside the horizon.
The Penrose diagram is shown in figure~7.

\vskip 0.3cm
\noindent
{\bf 3.1.1.7. $\omega=\mp\infty$}

\noindent
Taking the limit $\omega\rightarrow\mp\infty$
in eqs.\ (\ref{eq:13}) and (\ref{eq:14}) we obtain:
\begin{eqnarray}
   & & \phi= 0,  \label{eq:36a}  \\
   & & ds^2 = - \left( a^2 r^2 - ar \right) dt^2
              + \frac{dr^2}{a^2 r^2 - ar},
                \label{eq:36b}
\end{eqnarray}
with $a=2|\bar{\lambda}|$,
$\bar{\lambda}^2=\omega\lambda^2=\mbox{constant}<\infty$.
As one can easily see from eq.~(\ref{eq:15}),
the scalar curvature is constant, $R=-8\bar{\lambda}^2$.
Thus,
we obtain a black hole with a causal structure similar to the
$\omega=0$ case (see Fig.~2) \cite{lesa}.
By adding a Lagrangean matter term to the action (\ref{eq:1})
one recovers in this limit the $R=T$ theory.
In 4-dimensions the Einstein Theory of Relativity is
also recovered from the Brans-Dicke theory by taking
the limit $\omega\rightarrow +\infty$.
In this sense it is natural to consider the theory given by
eq.~(\ref{eq:1}) in the limit $\omega\rightarrow\mp\infty$
as the two-dimensional analog of General Relativity.
This is in contrast to the case $\omega=-\frac12$,
which is identical to vacuum planar General Relativity.

Since $\omega\rightarrow -\infty$
and $\omega\rightarrow +\infty$ have the same
solutions one can form a cyclic chain of diagrams,
the next one being the typical case $\omega=1$
(the diagrams with $\omega=-2$ and $\lambda^2 > 0$
or $\lambda^2<0$ -- case 3.3.1. and 3.3.2., respectively --,
can also enter in this cycle).

\vskip 0.3cm
\noindent
{\bf 3.1.2. $|A|<|B|$}

\noindent
This case yields naked singularities,
anti-de~Sitter spacetime and also black hole solutions.
Whenever $|A|<|B|$,
equation (\ref{eq:8}) takes the form,
\begin{equation}
  \phi = -\frac{1}{\omega+2} \ln
          \sinh \left( \sqrt{(\omega+2) \lambda^2} x \right)
         +\phi_0,          \label{eq:37}
\end{equation}
and the metric in the unitary gauge is given by
\begin{equation}
  ds^2 = - \coth^2 \left( \sqrt{(\omega+2)\lambda^2} x \right)
           \sinh^{ \frac{4\left(\omega+1\right)}{\omega+2}}
                   \left( \sqrt{(\omega+2)\lambda^2} x \right)
           dt^2
         + dx^2.                    \label{eq:38}
\end{equation}
To write eq.~(\ref{eq:38}) in the Schwarzschild gauge,
one defines the radial coordinate $r$,
\begin{equation}
  r=\frac{b^{\frac{\omega+1}{\omega+2}}}{a}
            \sinh^{\frac{2(\omega+1)}{\omega+2}}
                 \left(\sqrt{(\omega+2)\lambda^2} x \right),
                 \label{eq:39}
\end{equation}
where $a$ was defined in eq.~(\ref{eq:12})
and $b>0$.
Then the dilaton and metric fields take the form,
\begin{eqnarray}
   & & \phi= - \ln (ar)^{\frac{1}{2 \left( \omega+1 \right)}},
                \label{eq:41}  \\
   & & ds^2 = - \left[ a^2 r^2
              + b (ar)^{\frac{\omega}{\omega+1}} \right]
                dt^2
              + \frac{dr^2}{a^2 r^2
              + b (ar)^{\frac{\omega}{\omega+1}}},
                \label{eq:42}
\end{eqnarray}
where in eq.~(\ref{eq:41}) we have set to zero the
constant of integration.
As for case 3.1.1.\ we now divide case 3.1.2.\
in seven subsections according to the range of $\omega$.
In this section, as in 3.1.1., we choose $b=1$ ($\phi_0=0$).
Taking $b<0$ in eq.~(\ref{eq:14})
is equivalent to eq.~(\ref{eq:42}).

\vskip 0.3cm
\noindent
{\bf 3.1.2.1.  $0<\omega<+\infty$}

\noindent
For the typical case $\omega=1$ the metric takes the form
\begin{equation}
 ds^2 = - \left( a^2 r^2 + \sqrt{ar} \right) dt^2
        + \frac{dr^2}{a^2 r^2 + \sqrt{ar}}. \label{eq:43}
\end{equation}
The conformal radial coordinate can be found to be
\begin{equation}
 r_*=\frac{2}{3a} \ln \frac{\sqrt{ar}+1}{\sqrt{ar-\sqrt{ar}+1}}
    +\frac{2\sqrt3}{3a} \arctan \frac{2\sqrt ar-1}{\sqrt3}.
                 \label{eq:44}
\end{equation}
Defining the null coordinates $u=t-r_*$ and $v=t+r_*$,
we find in a Penrose diagram ($u,v$) that the singularity
and infinity are timelike.
Since there are no horizons,
the singularity is naked (see Fig.~8).

\vskip 0.3cm
\noindent
{\bf 3.1.2.2.  $\omega=0$}

\noindent
The metric is given by
\begin{equation}
  ds^2 = - \left( a^2r^2+1 \right) dt^2 + \frac{dr^2}{a^2r^2+1}.
                               \label{eq:45}
\end{equation}
Spacetime is non-singular and we have here
the two-dimensional anti-\-de~Sitter spacetime (see Fig.~9).
This is a solution of the Jackiw-Teitelboim theory.

\vskip 0.3cm
\noindent
{\bf 3.1.2.3. $-1<\omega<0$}

\noindent
A typical case is $\omega = -{1\over2}$,
which gives a solution of planar General Relativity.
The metric is
\begin{equation}
  ds^2 = - \left( a^2r^2 + \frac{1}{ar} \right) dt^2
         + \frac{dr^2}{a^2r^2 + \frac{1}{ar}}. \label{eq:46}
\end{equation}

The conformal coordinate is given by
\begin{equation}
 r_*= -\frac{1}{6a}
       \ln \left[ \frac{(1+ar)^2}{1-ar+ a^2r^2} \right]
      +\frac{1}{\sqrt3 a} \arctan \frac{2ar-1}{\sqrt3}.
                    \label{eq:47}
\end{equation}
In a Penrose diagram ($u,v$),
one can see that there is a timelike
naked singularity.
Infinity is also timelike just as
in case 3.1.2.1.\ (see Fig.~8).

\vskip 0.3cm
\noindent
{\bf 3.1.2.4. $\omega=-1$}

\noindent
This is the case provided by string theory.
The spacetime has been widely studied \cite{witt,dijk}.
In the Euclideanized version it is called the trumpet spacetime.
The Penrose diagram is identical to the 4-dimensional
Reissner-Nordstrom naked singularity (see Fig.~10).
Spacetime is asymptotically flat.

\vskip 0.3cm
\noindent
{\bf 3.1.2.5. $-2<\omega<-1$}

\noindent
For the typical case $\omega = -\frac{3}{2}$ the metric is
\begin{equation}
      ds^2 = - \left( a^2r^2 + a^3r^3 \right) dt^2
             + \frac{dr^2}{a^2r^2+a^3r^3}.  \label{eq:48}
\end{equation}
Performing the transformation $ar\rightarrow -ar$ we see
that this solution coincides with the black hole solution
given in eq.~(\ref{eq:33}) (see Fig.~5 for the Penrose
diagram).

For $\omega = -\frac43$ the metric is
\begin{equation}
      ds^2 = - \left( a^2r^2 + a^4r^4 \right) dt^2
             + \frac{dr^2}{a^2r^2+a^4r^4},  \label{eq:49}
\end{equation}
which corresponds to the naked singularity shown in
figure~11.

\vskip 0.3cm
\noindent
{\bf 3.1.2.6. $-\infty<\omega<-2$}

\noindent
As before, we consider a typical case $\omega=-3$.
Then the metric in Schwarz\-schild coordinates takes the form
\begin{equation}
   ds^2 = - \left[ a^2r^2 + (ar)^{3/2} \right] dt^2
          + \frac{dr^2}{a^2r^2 + (ar)^{3/2}}.  \label{eq:50}
\end{equation}
The Penrose diagram is shown in figure~12.
The singularity at $ar=0$ is represented by null lines.

\vskip 0.3cm
\noindent
{\bf 3.1.2.7. $\omega=\mp\infty$}

\noindent
In the limit $\omega\rightarrow\mp\infty$
equations (\ref{eq:41}) and (\ref{eq:42}) become:
\begin{eqnarray}
   & & \phi= 0,  \label{eq:50a}  \\
   & & ds^2 = - \left( a^2 r^2 + ar \right) dt^2
              + \frac{dr^2}{a^2 r^2 + ar}.
                \label{eq:50b}
\end{eqnarray}
This black hole solution coincides with the solution found in
section 3.1.1.7. (Fig.~2),
as it can be seen by redefining $r\rightarrow -r$.

\vskip 0.3cm
\noindent
{\bf 3.1.3. $A=|B|$}

\noindent
Whenever $A=|B|$, the solution of eq.~(\ref{eq:5})
takes the form of the linear dilaton
\begin{equation}
   \phi(x)=-\frac{\lambda^2}{\sqrt{(\omega+2)\lambda^2}}x
           + \phi_0.
                      \label{eq:51}
\end{equation}
The metric is given by
\begin{equation}
   ds^2=-e^{2ax}dt^2 + dx^2,   \label{eq:52}
\end{equation}
where the constant $a$ is defined in eq.~(\ref{eq:12}).
For $\omega=-1$ one obtains immediately from eq.~(\ref{eq:52})
that spacetime is of Minkowski type (see Fig.~13).
For $\omega\neq-1$ the dilaton and the metric
in the Schwarzschild gauge read
\begin{eqnarray}
 & & \phi=-\ln (ar)^{\frac{1}{2(\omega+1)}},
                               \label{eq:52a} \\
 & & ds^2 = -a^2r^2dt^2 + \frac{dr^2}{a^2 r^2}.
                               \label{eq:52b}
\end{eqnarray}
Taking $b=0$ in eq.~(\ref{eq:14}) is equivalent
to eq.~(\ref{eq:52b}).
This is not a  black hole solution with horizons at $ar=0$
($x=-\infty$), since there are no duplications.
The linear dilaton with $\omega=-2$ is treated in section 3.3.4.
The maximally extended spacetime is shown in figure~14.
Since the scalar curvature is constant, $R=-2a^2$, there are no
singularities.

\vskip 0.8cm
\noindent
{\bf 3.2. $(\omega+2)\lambda^2<0$}

\noindent
The solution of eq.~(\ref{eq:5}) changes from hyperbolic to
trigonometric and is given by
\begin{equation}
  \phi = -\frac{1}{\omega+2} \ln \left[
          C \cos \left( \sqrt{-(\omega+2)\lambda^2} x \right)
         +D \sin \left( \sqrt{-(\omega+2)\lambda^2} x \right)
                                 \right],
               \label{eq:53}
\end{equation}
where $C$ and $D$ are constants of integration.
This can be always put in the form
\begin{equation}
  \phi = -\frac{1}{\omega+2} \ln
          \cos \left( \sqrt{-(\omega+2)\lambda^2} x \right)
         +\phi_0,          \label{eq:53a}
\end{equation}
which is defined for
$|x|<\frac{\pi}{2} \frac{1}{\sqrt{-(\omega+2)\lambda^2}}$.
The metric in the unitary gauge can be written as
\begin{equation}
  ds^2 = - \tan^2 \left( \sqrt{-(\omega+2)\lambda^2} x \right)
           \cos^{ \frac{4\left(\omega+1\right)}{\omega+2}}
                   \left( \sqrt{-(\omega+2)\lambda^2} x \right)
           dt^2
         + dx^2.                    \label{eq:54}
\end{equation}
Schwarzschild coordinates are recovered if we write,
$\cos^2 \left( \sqrt{-(\omega+2) \lambda^2} x \right)
=(\bar{a}r)^{\frac{\omega+2}{\omega+1}} b^{-1}$,
where
$\bar{a}=
 \frac{2(\omega+1)\lambda^2}{\sqrt{-(\omega+2)\lambda^2}}$.
The metric is then:
\begin{equation}
       ds^2 = - \left[ b(\bar{a}r)^{\frac{\omega}{\omega+1}}
              - \bar{a}^2 r^2 \right]
                dt^2
              + \frac{dr^2}{b(\bar{a}r)^{\frac{\omega}{\omega+1}}
              - \bar{a}^2 r^2}.
                \label{eq:55}
\end{equation}

As before, we choose $b=1$ ($\phi_0=0$).
One can now work the maximal analytical extension.
We again divide into seven distinct cases.
It is easy to work out that the Penrose diagrams are
given by a 90 degrees rotation of the cases
3.1.1.1.-3.1.1.7.\ in section 3.1.1. (see Figs.~15-21).
We now quickly comment on each typical solution,
as given in section 3.1.1.

For $\omega=1$ there are two different diagrams
(see Fig.~15),
one with the singularity in the past,
the other with the singularity in the future.
This is in contrast with case 3.1.1.1.\
(see Fig.~1), where the two disconnected
pieces were simply replics of each other.
Region II is duplicated.
However, the horizons are cosmological rather than black
hole event horizons.

For $\omega=0$ this solution corresponds to the 2D
de~Sitter spacetime.
The maximal analytical extension is given in figure~16.
Region I is duplicated.
The diagram is rather similar to the one obtained by
pasting together diagrams of the four dimensional
de~Sitter spacetime.
Contrarily to the 4-dimensional case the Penrose
diagram is not observer dependent.

For $\omega=-1/2$ and $\omega=-1$ the timelike singularities
are naked (see Figs.~17 and 18, respectively).
Regions I of both diagrams are duplicated.
Infinities are spacelike and null for
$\omega=-1/2$ and $\omega=-1$, respectively.

For $\omega=-3/2$ and for $\omega=-4/3$ the singularities are
naked (see Fig.~19 and 20, respectively).

For $\omega=-3$ the Penrose diagram
looks like a Schwarzschild diagram,
where the singularities take the place of the Schwarzschild
infinities (see Fig.~21).

For $\omega=\mp\infty$ the Penrose diagram
is similar to the $\omega=0$ case (see Fig.~16).

\vskip 0.8cm
\noindent
{\bf 3.3. $(\omega+2)\lambda^2=0$}

\noindent
Here there are four distinct cases to be considered.

\vskip 0.3cm
\noindent
{\bf 3.3.1. $\lambda^2>0$ and $\omega=-2$}

\noindent
This solution gives a naked singularity.
In the unitary gauge, it is given by
\begin{eqnarray}
   & &  \phi=-\frac{\lambda^2}{2}x^2 + \phi_0, \\
   & &  ds^2 = - \lambda^2 x^2 e^{-2\lambda^2 x^2} dt^2+dx^2.
               \label{eq:56}
\end{eqnarray}
In Schwarzschild coordinates this gives the logarithmic case
mentioned earlier:
\begin{eqnarray}
   & &  e^{-2\phi}=-\frac{1}{2|\lambda|r}, \label{eq:56aa} \\
   & &  ds^2 = - 4\lambda^2r^2
                   \ln\left(-2|\lambda|r\right)^{-1} dt^2
               + \frac{dr^2}{4\lambda^2r^2
                   \ln\left(-2|\lambda|r\right)^{-1} },
          \label{eq:56a}
\end{eqnarray}
where we have set to zero in eq.~(\ref{eq:56aa}) the
constant of integration.
It is an interesting example.
The scalar curvature is
$R=12\lambda^2 +8 \lambda^2 \ln \left(-2|\lambda|r\right)$.
Thus, there are singularities at
$|\lambda|r=0$ and $|\lambda|r =-\infty$.
The Penrose diagram (see Fig.~22)
shows that the whole frontier is singular.
The solution can be called a highly naked singularity.

\vskip 0.3cm
\noindent
{\bf 3.3.2. $\lambda^2<0$ and $\omega=-2$}

\noindent
The Penrose diagram for this naked singularity
is given
by a 90 degrees rotation of the
previous case 3.3.1.\ (see Fig.~23).

\vskip 0.3cm
\noindent
{\bf 3.3.3. $\lambda^2=0$ and $\omega\ne -2$}

\noindent

It is straightforward to solve the differential equations
(\ref{eq:5})-(\ref{eq:6}) to give,
\begin{eqnarray}
  & & \phi=-\frac{1}{\omega+2} \ln x + \phi_0,
                        \label{eq:57} \\
  & & ds^2= -x^{\frac{2\omega}{\omega+2}} dt^2 + dx^2.
                        \label{eq:58}
\end{eqnarray}

In the Schwarzschild gauge the dilaton and metric read
\begin{eqnarray}
  & & \phi=-\frac{1}{2(\omega+1)} \ln c r, \label{eq:58a} \\
  & & ds^2=-(cr)^{\frac{\omega}{\omega+1}} dt^2,
           +\frac{dr^2}{(cr)^{\frac{\omega}{\omega+1}}}
                                           \label{eq:58b}
\end{eqnarray}
where $c=\frac{2(\omega+1)}{\omega+2}$;
we have in eq.~(\ref{eq:58a}) set to zero the constant of
integration.
For $\omega=-1$ one has to redefine the radial coordinate $r$
as in section 3.1.1.4.

{}From the scalar curvature,
$R=\frac{4\omega}{(\omega+2)^2}
(cr)^{-\frac{\omega+2}{\omega+1}}$,
we conclude that for all values of $\omega$
(except $\omega=0$ and $\omega=\mp\infty$)
there are singularities;
since horizons are absent,
these singularities are naked.
For $-\infty<\omega<-2$ the
singularities are null (see Fig.~12),
and for other values of $\omega$ the singularities
are spacelike and timelike (see Figs.~10,11,24,25).
For $\omega=0$ the spacetime is of
the Minkowski type (see Fig.~13).
In the limit $\omega\rightarrow\mp\infty$ the metric
(\ref{eq:58}) goes to the Rindler metric,
whose maximal analitical extension yields the
Minkowski spacetime.

\vskip 0.3cm
\noindent
{\bf 3.3.4. $\lambda^2=0$ and $\omega=-2$}

\noindent
The solution is
\begin{eqnarray}
  & & \phi= \alpha x + \phi_0, \\
  & &  ds^2=-e^{4\alpha x} dt^2 + dx^2, \label{eq:59}
\end{eqnarray}
where $\alpha$ is any constant with the
same units as $\lambda$.
This case yields the linear dilaton solution of $\omega=-2$,
which is identical to the case analised in section~3.1.3.
The corresponding Penrose diagram for $\alpha\neq 0$
is given in figure~14.
$\alpha=0$ gives the Minkowski spacetime.

\vskip 1cm

\noindent
{\bf 4. The ADM Masses of the Solutions}

\vskip 3mm

\noindent
A useful and important quantity that appears in the
theory of General Relativity is the
Arnowitt-Deser-Misner (ADM) total mass,
which can be defined for isolated black holes.
For the two-dimensional black holes one can calculate
an analogous quantity.
In a static spacetime there is a timelike Killing vector,
$\xi^a=\left( \frac{\partial}{\partial t} \right)^a$,
which implies the existence of a conserved quantity given by
$p^0={T^0}_b \xi^b$,
where ${T^a}_b$ is given in eq.~(\ref{eq:2}).
The corresponding charge density is a total divergence,
$P^0=\sqrt{-g}p^0$.
The total charge is then the analogue of the ADM mass,
which can be found through the equation
\begin{equation}
M_{BH}=\int_{\infty} P^0 d\rho
      =\int_{\infty} \sqrt{-g} {T^0}_a \xi^a d\rho,
       \label{eq:60}
\end{equation}
where $\rho$ represents the spatial coordinate
at spatial infinity.

In this paper we have worked with the unitary
and Schwarzschild gauges.
In order to compare the expression for the ADM masses
we calculate the integral of eq.~(\ref{eq:60}) in both gauges.

The idea is to subtract in a correct manner the black hole from
the corresponding linear dilaton solution at spatial infinity.
In the unitary gauge the linear dilaton solution is given by
\begin{eqnarray}
  & & \bar{\phi}=-\frac{\lambda^2}{\sqrt{(\omega+2)\lambda^2}}x,
                    \label{eq:61} \\
  & & ds^2 = \bar{g}_{ab} dx^a dx^b
           = -e^{2ax} dt^2+ dx^2, \label{eq:62}
\end{eqnarray}
where
$a=\frac{2(\omega+1)\lambda^2}{\sqrt{(\omega+2)\lambda^2}}$
(see equations (\ref{eq:51}) and (\ref{eq:52})).
At spatial infinity, $x\rightarrow\infty$, the black hole
solution (\ref{eq:9})-(\ref{eq:10}) can be written as
\begin{eqnarray}
  & & \phi=\bar{\phi}+\varphi,   \label{eq:63}  \\
  & & g_{ab}=\bar{g}_{ab}+h_{ab},  \label{eq:64}
\end{eqnarray}
where $\bar{\phi}$ and $\bar{g}_{ab}$ are the background
(linear) dilaton solutions given in equations
(\ref{eq:61})-(\ref{eq:62}),
and $\varphi$ and $h_{ab}$ are the fluctuations above
the background due to the presence of the black hole.
Now, when $x\rightarrow\infty$, $\sqrt{-g}{T_0}^0$ goes into
\begin{equation}
 \sqrt{-g}{T_0}^0=
   \frac{\partial}{\partial x}
   \left[ e^{2\sqrt{(\omega+2)\lambda^2}x}
          \left(  \frac{\partial \varphi}{\partial x}
          +\frac{\lambda^2}{\sqrt{(\omega+2)\lambda^2}} h_{11}
          \right)  \right].      \label{eq:65}
\end{equation}
{}From (\ref{eq:9})-(\ref{eq:10}) and (\ref{eq:61})-(\ref{eq:64})
we find
$\frac{\partial \varphi}{\partial x}
=\frac12 \frac{\lambda^2}{\sqrt{(\omega+2)\lambda^2}}
e^{-2(\omega+2)\phi_0}
e^{-2\sqrt{(\omega+2)\lambda^2}x}$
and $h_{11}=0$.
Thus, from (\ref{eq:60}) and (\ref{eq:65}) one obtains
\begin{equation}
 M_{BH}=\frac12 \frac{\lambda^2}{\sqrt{(\omega+2)\lambda^2}}
        e^{-2(\omega+2)\phi_0}.
       \label{eq:66}
\end{equation}
In the unitary gauge the mass depends on $\phi_0$,
the value of $\phi$ at the horizon $x=0$ (see also \cite{witt}).
Following the same procedure for the Schwarzschild gauge we find
from eqs.\ (\ref{eq:13})-(\ref{eq:14}) and the corresponding
linear dilaton solutions (\ref{eq:52a})-(\ref{eq:52b})
that the total divergence is given asymptotically by
\begin{equation}
 \sqrt{-g}{T_0}^0=
   \frac{\partial}{\partial r}
   \left[ (ar)^{\frac{2\omega+3}{\omega+1}}
          a \frac{\partial \varphi}{\partial r}
          + \frac14 \frac{a}{\omega+1}
            (ar)^{\frac{3\omega+4}{\omega+1}} h_{11}
   \right].      \label{eq:67}
\end{equation}
One can easily find that in this case $\varphi=0$ and
$h_{11}=b (ar)^{-\frac{3\omega+4}{\omega+1}}$.
Thus, from (\ref{eq:60}) one obtains
\begin{equation}
 M_{BH}=\frac12 \frac{\lambda^2}{\sqrt{(\omega+2)\lambda^2}} b.
       \label{eq:68}
\end{equation}
In the Schwarzschild gauge the mass of the black hole is
related to $b$,
the value of $ar$ at the horizon,
since there $b= (ar)^{\frac{\omega+2}{\omega+1}}$.
This is what one should expect from experience with the
Schwarzschild metric.
Through eqs.\ (\ref{eq:66}) and (\ref{eq:68}) we can relate
$\phi_0$ and $b$ by
\begin{equation}
  b=e^{-2(\omega+2)\phi_0}. \label{eq:69}
\end{equation}
Note that expression (\ref{eq:66}) applies only for
the solutions of section 3.1.1.,
where $A>|B|$ ($b>0$).
On the other hand,
eq.~(\ref{eq:68}) is valid for $b>0$, $b<0$ and $b=0$.

It is now interesting to analise expression (\ref{eq:68})
for the several solutions we have obtained.

When $(\omega+2)\lambda^2>0$, $\lambda^2>0$ and $b>0$ one
has seven different black holes
(3.1.1.1.-3.1.1.5.\ and 3.1.1.7, Figs.~1-5,7, respectively),
and the ADM mass is non-negative, $M_{BH}\geq 0$.
For $\omega=0$ one obtains the mass of the
Teitelboim-Jackiw black hole,
$M(\omega=0)=\frac{1}{2} \sqrt{\frac12} b |\lambda|$.
For $\omega=-\frac12$ one obtains the two-dimensional mass
(in this case a surface density)
of the corresponding planar black hole in General Relativity,
$M(\omega=-\frac12)=\frac12 \sqrt{\frac23} b |\lambda|$.
For $\omega=-1$ one obtains Witten's value,
$M(\omega=-1)=\frac12 b |\lambda|$.
In the limit $\omega\rightarrow\mp\infty$ one obtains
the surprising result
$M(\omega=\mp\infty)=0$,
although from the causal structure of the solution
3.1.1.7.\ one can ascertain the presence of a black hole.
When $(\omega+2)\lambda^2>0$, $\lambda^2>0$ and $b<0$,
which corresponds to cases 3.1.2.1.-3.1.2.5., 3.1.2.7.\
(see Figs.~2,5,8-11),
one has solutions with non-positive mass,
some are naked singularities.
When $(\omega+2)\lambda^2>0$ and $b=0$ one has the linear dilaton,
and by construction these solutions have zero mass.

When $(\omega+2)\lambda^2>0$, $\lambda^2<0$ and $b>0$,
which is the case 3.1.1.6.\ (Fig.~6),
one has a black hole with negative mass.
The corresponding naked singularity,
with $b<0$ (3.1.2.6., Fig.~12),
has positive mass.

When $(\omega+2)\lambda^2=0$ and $\lambda^2>0$
(3.3.1., Fig.~22) one has a naked singularity with infinite mass,
which is a highly naked singularity.
The case $\lambda^2<0$ (3.3.2, Fig.~23) gives a highly naked
singularity with infinite negative mass.
For $\lambda^2=0$ (3.3.3.\ and 3.3.4., Figs.~10-13,24)
the ADM mass is zero.
Some of these are (massless) naked singularities.

When $(\omega+2)\lambda^2<0$ one has always imaginary masses
(see section~3.2., Figs.~15-21).
The ADM mass concept in these cases is meaningless.

\vskip 1cm

\noindent
{\bf 5. Conclusions}

\vskip 3mm

\noindent
We have presented a bewildering variety of solutions which
appear in a theory given by action (\ref{eq:1})
-- see table~1 for a summary.
The theory has two parameters, the cosmological constant
$\lambda$ and the Brans-Dicke parameter $\omega$.
For special values of $\omega$ some important cases arise.
Thus, (i) $\omega\rightarrow\mp\infty$ yields a purely geometric
          two-dimensional theory,
      (ii) $\omega=0$ gives the constant curvature theory,
      (iii) $\omega=-\frac12$ is equivalent to (vacuum) planar
            General Relativity, and
      (iv) $\omega=-1$ is the action one obtains in string
           theory by imposing conformal invariance on the
           string world sheet.

Out of all the solutions maybe the most interesting are the
black hole solutions with real ADM masses.
Depending on the value of $\omega$ and $\lambda^2$ they
have several different causal structures.
There are structures similar to
(i) the extreme Reissner-Nordstrom
    ($0<\omega<+\infty$, $\lambda^2>0$ and $A>|B|$),
(ii) the non-singular black holes similar to the exact
     string theory
     ($\omega=0$, $\lambda^2>0$ and $A>|B|$;
     $\omega=\mp\infty$, $\omega\lambda^2>0$),
(iii) the Schwarzschild black hole analogue
      ($\omega=-1$, $\lambda^2>0$ and $A>|B|$).
There are also new structures,
(i) $-1<\omega<0$, $\lambda^2>0$ and $A>|B|$
    being a Schwarzschild-like black hole with a
    timelike infinity,
(ii) $-2<\omega<-1$, $\lambda^2>0$ and $A>|B|$
     yielding two Penrose diagrams, one of them tiles
     the whole plane, and
(iii) $\omega=-2$ and $\lambda^2\neq0$
      giving the highly naked singularity.
One also finds several naked singularities, obtained by rotating
through $90^{\circ}$ the above mentioned diagrams, as well as
many other solutions giving naked singularities, anti-de~Sitter,
de~Sitter and Minkowski spacetimes.
Having studied the geometrical structure of this quite general
two-dimensional theory,
one can now explore its physical consequences,
such as thermodynamical properties and black hole
evaporation in a fashion similar to what has been done recently
\cite{strom}

\vskip 1cm

\noindent
{\bf Acknowledgements}

\noindent
JPSL acknowledges grants from JNICT (Portugal) and CNPq (Brazil).
PMS acknowledges a grant from JNICT (Portugal).

\newpage

{\bf Table Caption}

\vskip 1cm

{\bf Table 1.}  The table summarizes all the cases discussed
                in the text.
                In the classification column the short
                names used are:
                wS - with singularity,
                woS - without singularity,
                BH - black hole,
                NS - naked singularity, and
                LD - linear dilaton.
\vskip 1.5cm

{\bf Figure Captions}

\vskip 1cm

{\bf Figure 1.} Penrose diagram for the black hole with $\omega=1$,
                $\lambda^2>0$ and $A>|B|$
                (representative case of $0<\omega<+\infty$).
                Region I corresponds to two triangles glued at the
                horizon.
                Double lines represent spacetime singularities and
                simple lines infinities and horizons.
                This diagram is similar to the extreme
                four-dimensional
                Reissner-Nordstrom black hole diagram.

\vskip 5mm

{\bf Figure 2.} Penrose diagram for
                (i) the maximally extended black
                hole of the Teitelboim-Jackiw theory, $\omega=0$,
                $\lambda^2>0$ and $A>|B|$, and
                (ii) the black hole with $\omega=\mp\infty$,
                $\omega\lambda^2>0$ and
                $A\neq|B|$.
                Region I is also duplicated.
                There is an infinite chain of regions,
                none of them contains a singularity.

\vskip 5mm

{\bf Figure 3.} Penrose diagram for the black hole
                with $\omega=-\frac12$
                (representative case of $-1<\omega<0$),
                $\lambda^2>0$ and $A>|B|$.
                Regions I and I' describe two identical
                but space-inverted regions.
                Regions II and II' describe two identical but
                time-reversed regions,
                the black hole and the white hole,
                respectively.
                Infinities are timelike lines.
                The case $\omega=-\frac12$ is also a black hole in
                General Relativity.

\vskip 5mm

{\bf Figure 4.} Penrose diagram for the black hole with
                $\omega=-1$, $\lambda^2>0$ and $A>|B|$.
                It is similar to the diagram of figure~3,
                but now infinities are null, i.e., is identical to
                the Schwarzschild diagram.
                This is the black hole of string theory.

\vskip 5mm

{\bf Figure 5.} Penrose diagram for the black hole
                with $\omega=-\frac32$
                (representative case of $-2<\omega<-1$),
                $\lambda^2>0$ and $A>|B|$.
                There are an infinite chain of universes containing
                spacelike and timelike singularities.

\vskip 5mm

{\bf Figure 6.} Penrose diagram for the black hole
                with $\omega=-\frac43$
                (also representative of $-2<\omega<-1$),
                $\lambda^2>0$ and $A>|B|$.
                There are an infinite chain of universes containing
                spacelike singularities.
                The diagram tiles the whole plane.

\vskip 5mm

{\bf Figure 7.} Penrose diagram for the black hole
                with $\omega=-3$
                (representative case of $-\infty<\omega<-2$),
                $\lambda^2<0$ and $A>|B|$.
                The singularities are null and the
                infinities are timelike.

\vskip 5mm

{\bf Figure 8.} Penrose diagram for the naked singularity
                in the cases
                (i) $0<\omega<+\infty$, $\lambda^2>0$ and $|A|<|B|$,
                (ii) $-1<\omega<0$, $\lambda^2>0$ and $|A|<|B|$.
                The two points represent timelike infinity.
                Spacelike and null infinities are represented
                by timelike lines.

\vskip 5mm

{\bf Figure 9.} Penrose diagram for the anti-de~Sitter
                spacetime with
                $\omega=0$, $\lambda^2>0$ and and $|A|<|B|$.

\vskip 5mm

{\bf Figure 10.} Penrose diagram for the naked timelike
                 singularity in the cases
                 (i) $\omega=-1$, $\lambda^2>0$ and $|A|<|B|$,
                 (ii) $\omega=-1$ and $\lambda^2=0$
                 (in this case $\lambda$ must be dropped out).

\vskip 5mm

{\bf Figure 11.} Penrose diagram for the naked timelike
                 singularity in the cases
                 (i) $\omega=-\frac43$, $\lambda^2>0$ and $|A|<|B|$,
                 (ii) $\omega=-\frac43$ and $\lambda^2=0$
                 (in this case $a$ must be replaced by $c$).

\vskip 5mm

{\bf Figure 12.} Penrose diagram for the naked null singularity
                 in the cases
                 (i) $-\infty<\omega<-2$, $\lambda^2<0$ and $|A|<|B|$,,
                 (ii) $-\infty<\omega<-2$ and $\lambda^2=0$
                 (in this case $a$ must be replaced by $c$).

\vskip 5mm

{\bf Figure 13.} Penrose diagram for the Minkowski spacetime,
                 corresponding to cases
                 (i) $\omega=-1$, $\lambda^2>0$ and $A=|B|$,
                 (ii) $\omega=\mp\infty$ and $\lambda^2=0$,
                 (iii) $\omega=0$ and $\lambda^2=0$,
                 (iv) $\lambda^2=0$, $\omega=-2$ and $\alpha=0$.

\vskip 5mm

{\bf Figure 14.} Penrose diagram for the linear dilaton ,
                 appearing in the cases
                 (i) $(\omega+2)\lambda^2>0$
                 ($\omega\neq-1$) and $A=|B|$,
                 (ii) $\lambda^2=0$, $\omega=-2$ and
                 $\alpha\neq 0$.

\vskip 5mm

{\bf Figure 15.} Penrose diagrams for the cosmological solutions with
                 $0<\omega<+\infty$ and $\lambda^2<0$.
                 There are two disconnected pieces,
                 one having the singularity in the past,
                 the other with the singularity in the future.
                 This diagram can be obtained from the diagram of
                 Fig.~1 by a 90 degrees rotation.

\vskip 5mm

{\bf Figure 16.} Penrose diagram for the de~Sitter spacetime
                 with
                 (i) $\omega=0$ and $\lambda^2<0$, and
                 (ii) $\omega=\pm\infty$, $\omega\lambda^2<0$.
                 This diagram is constructed by horizontally
                 pasting diagrams of the
                 4-dimensional de~Sitter spacetime.
                 This diagram can be obtained from the
                 diagram of Fig.~2 by a 90 degrees rotation.

\vskip 5mm

{\bf Figure 17.} Penrose diagram for the naked singularity
                 with $-1<\omega<0$ and $\lambda^2<0$.
                 This diagram can be obtained from the diagram of
                 Fig.~3 by a 90 degrees rotation.

\vskip 5mm

{\bf Figure 18.} Penrose diagram for the naked singularity
                 with $\omega=-1$ and $\lambda^2<0$.
                 This diagram can be obtained from the diagram of
                 Fig.~4 by a 90 degrees rotation.

\vskip 5mm

{\bf Figure 19.} Penrose diagram for the naked singularity
                 with $\omega=-\frac32$ and $\lambda^2<0$.
                 Singularities are timelike and spacelike.
                 This diagram can be obtained from the
                 diagram of Fig.~5 by a 90 degrees rotation.

\vskip 5mm

{\bf Figure 20.} Penrose diagram for the naked singularity
                 with $\omega=-\frac43$ and $\lambda^2<0$.
                 Singularities are timelike.
                 The diagram covers the whole plane.
                 This diagram can be obtained from the
                 diagram of Fig.~6 by a 90 degrees rotation.

\vskip 5mm

{\bf Figure 21.} Penrose diagram for the naked singularity
                 with $-\infty<\omega<-2$ and $\lambda^2>0$.
                 This diagram can be obtained from the diagram of
                 Fig.~7 by a 90 degrees rotation.

\vskip 5mm

{\bf Figure 22.} Penrose diagram for the highly naked singularity
                 with $\omega=-2$ and $\lambda^2>0$.
                 The whole frontier of the diagram is singular.

\vskip 5mm

{\bf Figure 23.} Penrose diagram for the highly naked singularity
                 with $\omega=-2$ and $\lambda^2<0$.
                 The whole frontier of the diagram is singular.
                 This diagram can be obtained from the diagram of
                 Fig.~22 by a 90 degrees rotation.

\vskip 5mm

{\bf Figure 24.} Penrose diagram for the naked singularity in the
                 cases
                 (i) $0<\omega<+\infty$ and $\lambda^2=0$,
                 (ii) $-1<\omega<0$ and $\lambda^2=0$.
                 Singularities are timelike.

\vskip 5mm

{\bf Figure 25.} Penrose diagram for the naked singularity in the
                 case $\omega=-\frac32$ and $\lambda^2=0$.
                 Singularities are timelike and spacelike.
                 The whole frontier is singular.

\newpage

\noindent
{\bf Table 1}

\vskip 1cm

{\scriptsize

\noindent
\begin{tabular}{|l|l|l|l|l|l|}                  \hline
Main                                            &
Subsidiary                                      &
Range of $\omega$                               &
Section in                                      &
Figure                                          &
Classification                                  \\
division                                        &
division                                        &
                                                &
the text                                        &
number                                          &
                                                \\ \hline
                                                &
                                                &
$0<\omega<+\infty$                              &
3.1.1.1.                                        &
Fig.~1                                          &
BH wS (Reissner-Nordstrom type)                 \\ \cline{3-6}
                                                &
                                                &
$\omega=0$                                      &
3.1.1.2.                                        &
Fig.~2                                          &
Black Hole woS                                  \\ \cline{3-6}
                                                &
                                                &
$-1<\omega<0$                                   &
3.1.1.3.                                        &
Fig.~3                                          &
Black Hole wS                                   \\ \cline{3-6}
                                                &
$A>|B|$                                         &
$\omega=-1$                                     &
3.1.1.4.                                        &
Fig.~4                                          &
BH wS (Schwarzschild type)                      \\ \cline{3-6}
                                                &
($b>0$)                                         &
$-2<\omega<-1$                                  &
3.1.1.5.                                        &
Figs.~5,6                                               &
Black Holes wS                                  \\ \cline{3-6}
                                                &
                                                &
$-\infty<\omega<-2$                             &
3.1.1.6.                                        &
Fig.~7                                          &
Black Hole wS                                   \\ \cline{3-6}
                                                &
                                                &
$\omega=\mp\infty$                              &
3.1.1.7.                                        &
Fig.~2                                          &
Black Hole woS                                  \\ \cline{2-6}
                                                &
                                                &
$0<\omega<+\infty$                              &
3.1.2.1.                                        &
Fig.~8                                          &
Naked singularity                               \\ \cline{3-6}
                                                &
                                                &
$\omega=0$                                      &
3.1.2.2.                                        &
Fig.~9                                          &
Anti-de~Sitter spacetime                        \\ \cline{3-6}
                                                &
                                                &
$-1<\omega<0$                                   &
3.1.2.3.                                        &
Fig.~8                                          &
Naked singularity                               \\ \cline{3-6}
$(\omega+2)\lambda^2>0$                         &
$|A|<|B|$                                       &
$\omega=-1$                                     &
3.1.2.4.                                        &
Fig.~10                                         &
Naked singularity                               \\ \cline{3-6}
                                                &
($b<0$)                                         &
$-2<\omega<-1$                                  &
3.1.2.5.                                        &
Fig.~11,5                                               &
Naked singularity and BH ws                     \\ \cline{3-6}
                                                &
                                                &
$-\infty<\omega<-2$                             &
3.1.2.6.                                        &
Fig.~12                                         &
Naked singularity                               \\ \cline{3-6}
                                                &
                                                &
$\omega=\mp\infty$                              &
3.1.2.7.                                        &
Fig.~2                                          &
Black Hole woS                                  \\ \cline{2-6}
                                                &
$A=|B|$                                         &
$\omega=-1$                                     &
3.1.3.                                          &
Fig.~13                                         &
Minkowski spacetime (LD)                        \\ \cline{3-6}
                                                &
($b=0$)                                         &
$\omega\neq-1$                                  &
3.1.3.                                          &
Fig.~14                                         &
Anti-de~Sitter spacetime (LD)                   \\ \hline
                                                &
                                                &
$0<\omega<+\infty$                              &
3.2.                                            &
Fig.~15                                         &
Future and past spacelike singularities         \\ \cline{3-6}
                                                &
                                                &
$\omega=0$                                      &
3.2.                                            &
Fig.~16                                         &
de~Sitter spacetime                             \\ \cline{3-6}
                                                &
                                                &
$-1<\omega<0$                                   &
3.2.                                            &
Fig.~17                                         &
Naked singularity                               \\ \cline{3-6}
$(\omega+2)\lambda^2<0$                         &
                                                &
$\omega=-1$                                     &
3.2.                                            &
Fig.~18                                         &
Naked singularity                               \\ \cline{3-6}
                                                &
                                                &
$-2<\omega<-1$                                  &
3.2.                                            &
Fig.~19,20                                      &
Naked singularities                             \\ \cline{3-6}
                                                &
                                                &
$-\infty<\omega<-2$                             &
3.2.                                            &
Fig.~21                                         &
Naked singularity                               \\ \cline{3-6}
                                                &
                                                &
$\omega=\mp\infty$                              &
3.2.                                            &
Fig.~16                                         &
de~Sitter spacetime                             \\ \hline
                                                &
$\lambda^2>0$                                   &
$\omega=-2$                                     &
3.3.1                                           &
Fig.~22                                         &
Naked singularity                               \\ \cline{2-6}
                                                &
$\lambda^2<0$                                   &
$\omega=-2$                                     &
3.3.2                                           &
Fig.~23                                         &
Naked singularity                               \\ \cline{2-6}
                                                &
                                                &
$0<\omega<+\infty$                              &
3.3.3.                                          &
Fig.~24                                         &
Naked singularity                               \\ \cline{3-6}
                                                &
                                                &
$\omega=0$                                      &
3.3.3.                                          &
Fig.~13                                         &
Minkowski spacetime                             \\ \cline{3-6}
                                                &
                                                &
$-1<\omega<0$                                   &
3.3.3.                                          &
Fig.~24                                         &
Naked singularity                               \\ \cline{3-6}
$(\omega+2)\lambda^2=0$                         &
$\lambda^2=0$                                   &
$\omega=-1$                                     &
3.3.3.                                          &
Fig.~10                                         &
Naked singularity                               \\ \cline{3-6}
                                                &
                                                &
$-2<\omega<-1$                                  &
3.3.3.                                          &
Fig.~11,25                                              &
Naked singularities                             \\ \cline{3-6}
                                                &
                                                &
$-\infty<\omega<-2$                             &
3.3.3.                                          &
Fig.~12                                         &
Naked singularity                               \\ \cline{3-6}
                                                &
                                                &
$\omega=\mp\infty$                              &
3.3.3.                                          &
Fig.~13                                         &
Minkowski spacetime                             \\ \cline{2-6}
                                                &
$\lambda^2=0$                                   &
$\omega=-2$ ($\alpha=0$)                        &
3.3.4.                                          &
Fig.~13                                         &
Minkowski spacetime (LD)                        \\ \cline{3-6}
                                                &
                                                &
$\omega=-2$ ($\alpha\neq0$)                     &
3.3.4.                                          &
Fig.~14                                         &
anti-de~Sitter (LD)                             \\ \hline
\end{tabular}
}

\end{document}